\documentclass[%
 aip,
rsi,%
 amsmath,amssymb,
preprint,%
]{revtex4-1}

\usepackage{graphicx}
\usepackage{dcolumn}
\usepackage{bm}
\usepackage{xcolor}
\usepackage{amsmath}
\usepackage{siunitx}
\usepackage{sidecap}
\usepackage{placeins}

\newcommand{\beq}{\begin{eqnarray}}
\newcommand{\eeq}{\end{eqnarray}}

\begin{document}

\title{A drift kinetic model for the expander region of a magnetic mirror}

\author{B. A. Wetherton}
\email{wetherton@lanl.gov}
\affiliation{Los Alamos National Laboratory, Los Alamos, New Mexico 87545, USA}
\author{A. Le}
\affiliation{Los Alamos National Laboratory, Los Alamos, New Mexico 87545, USA}
\author{J. Egedal}%
\affiliation{University of Wisconsin, Madison, Wisconsin 53706, USA}
\author{C. Forest}%
\affiliation{University of Wisconsin, Madison, Wisconsin 53706, USA}
\author{W. Daughton}
\affiliation{Los Alamos National Laboratory, Los Alamos, New Mexico 87545, USA}
\author{A. Stanier}
\affiliation{Los Alamos National Laboratory, Los Alamos, New Mexico 87545, USA}
\author{S. Boldyrev}%
\affiliation{University of Wisconsin, Madison, Wisconsin 53706, USA}
\affiliation{Center for Space Plasma Physics, Space Science Institute, Boulder, CO 80301, USA}

\date{\today}

\begin{abstract}
We present a drift kinetic model for the free expansion of a thermal plasma out of a magnetic nozzle. This problem relates to plasma space propulsion systems, natural environments such as the solar wind, and end losses from the expander region of mirror magnetically confined fusion concepts such as the Gas Dynamic Trap. The model incorporates trapped and passing orbit types encountered in the mirror expander geometry and maps to an upstream thermal distribution. This boundary condition and quasineutrality require the generation of an ambipolar potential drop of $\sim5 T_e/e$, forming a thermal barrier for the electrons. The model for the electron and ion velocity distributions and fluid moments is confirmed with data from a fully kinetic simulation. Finally, the model is extended to account for a population of fast sloshing ions arising from neutral beam heating within a magnetic mirror, again resulting in good agreement with a corresponding kinetic simulation.
\end{abstract}

\maketitle
\section{Introduction} \label{sec:intro}
The axisymmetric magnetic mirror was one of the first fusion reactor concepts \cite{Post1958,Post1973}, in part due to its simple geometry relative to the Tokamaks and stellarators that dominate the conversation in magnetically-confined fusion today. Initial interest in magnetic mirrors was largely doused by poor electron confinement due to large end losses of nearly-parallel travelling particles. Additionally, the simple mirror is unstable to curvature-driven magnetohydrodynamic (MHD) interchange modes, though this can be mitigated through several approaches. 

One approach of particular interest is the Gas Dynamic Trap (GDT) of Mirnov and Ryutov \cite{Mirnov1979}, which relies on the concept of pressure-weighted curvature to eliminate MHD interchange modes and consists of mirror coils separated by a solenoid long enough and with plasma density high enough to ensure that trapped ions are collisional and the outflows into the expanders are supersonic. The GDT experiment in Novosibirsk \cite{Ivanov2013} has demonstrated axisymmetric MHD stability \cite{Ivanov1994,Anikeev1997}, confinement of high energy sloshing ions sourced by a neutral beam, thermal confinement of electrons up to a temperature of \SI{940}{\electronvolt} via an ambipolar potential drop with average ion energies on the order of \SI{10}{\kilo \electronvolt} \cite{Bagryansky2014,Bagryansky2015a,Bagryansky2015}. These results show that the confinement is controlled by parallel processes with cross-field transport being negligible. The sloshing ions help to eliminate kinetic instabilities from the GDT configuration \cite{Kesner1980}. With the advent of High Temperature Superconducting (HTS) coils, magnetic mirrors can achieve mirror ratios far beyond what was available to early mirror experiments, greatly improving confinement. This, in tandem with promising results from GDT, brings us to revisit the axisymmetric magnetic mirror. 

In this paper, we examine the kinetic physics leading to parallel losses out of the expanding magnetic field at the end of a GDT. While the central cell of a GDT is by design long compared to the typical Coulomb collisional mean-free paths, the expander region is shorter and practically collisionless. For our analysis, we adopt a drift-kinetic treatment of the collisionless expansion of a thermal magnetized plasma along the axis of an expanding magnetic nozzle. We focus on the parallel transport processes that dominate in open field line geometries, and we do not consider cross-field transport. Coming from a space propulsion context, a drift-kinetic theoretical treatment of this process is given by Martinez-Sanchez et al. \cite{Martinez-Sanchez2015}, where the ions are treated as a monoenergetic population. There, the ambipolar potential along the magnetic axis is solved for by demanding quasineutrality and accounting for the boundary conditions. Skovorodin \cite{Skovorodin2019InfluenceTrap} has made a more complete theoretical treatment in application to the magnetic mirror where ions are also thermal. Here, we similarly include a thermal ion population, and expand to allow for a population of fast sloshing ions \cite{kotelnikov1985}. A similar treatment also forms the basis for a model of the global temperature profile of the solar wind \cite{Boldyrev2020ElectronWind}. That solar wind model focuses on regions far from the thermal source (at the solar corona), and it is dominated by collisionally scattered electrons that are trapped by a slowly-varying electric potential. In this paper, we consider the details of the region near the source where the profiles of the electric and magnetic fields, as well as the plasma density and temperature, vary relatively rapidly.

In Section \ref{sec:model} we introduce the drift-kinetic model and its application to the GDT expander geometry. The model is applied to a device with large mirror ratio and realistic deuterium mass ratio in Section \ref{sec:potential}, showing the generation of an ambipolar potential that acts as a natural thermal barrier to electrons. Section \ref{sec:basesim} verifies the model with results from a fully-kinetic particle-in-cell (PIC) simulation using the code VPIC \cite{Bowers2008} approximating the GDT expander region. Section \ref{sec:slosh} introduces another VPIC simulation, which is identical to the first save for the addition of a sloshing ion beam. The beam induces additional local trapping effects that are properly captured in the guiding center model. We conclude the paper in Section \ref{sec:conc}.

\section{Drift-kinetic model for expander physics}\label{sec:model}
The generation of an ambipolar potential that helps to confine hot electrons is one of the most striking aspects of the GDT experiment. In this section, we will explain how this potential is generated by examining the kinetic forms of the particle distribution functions in the magnetic mirror expander geometry. We start by examining the form of the distribution functions given a profile for the magnetic field strength and ambipolar potential.

The fundamental kinetic physics of the mirror expander region can be captured through a drift-kinetic, guiding center model. The mirror field is assumed to be strong enough for all particles to remain well-magnetized, and as such, particles conserve their adiabatic invariant magnetic moment $\mu = m v_\perp^2/2B$. To zeroth order, the guiding center orbits of well-magnetized particles follow magnetic field lines. We apply a single flux tube model similar to that of Egedal et al. \cite{Egedal2013, egedal:2008jgr} that maps the distribution function at a point along the flux tube to the distribution of an upstream population feeding the flux tube through Liouville's theorem in a collisionless plasma. The fundamental difference in the model presented here is the application of conditions expected to be encountered in the magnetic mirror expander. The model bears many similarities to the model of Skovorodin \cite{Skovorodin2019InfluenceTrap}, with the primary difference being where the upstream distribution is sourced as Maxwellian (internal to the device for our model, at the mirror throat for Skovorodin). 

\begin{figure}
	\centering
    \includegraphics[width=1.0\linewidth]{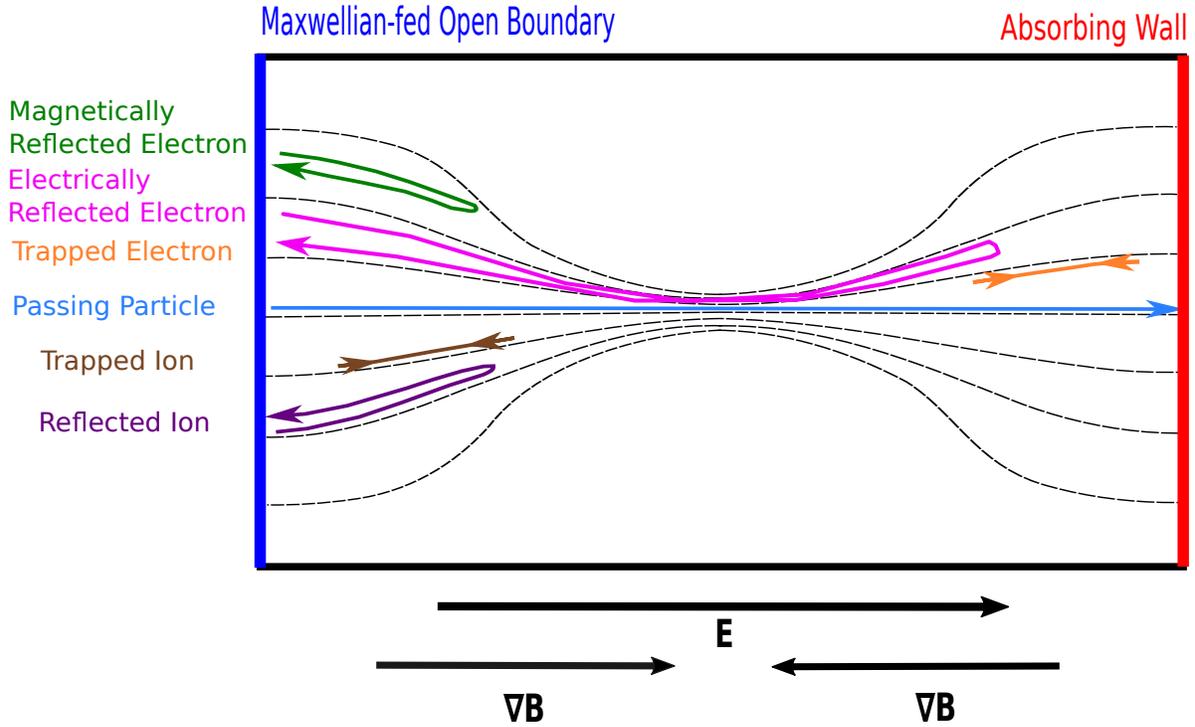}
	\caption{An illustration of the GDT expander simulation/model setup with selected magnetic field lines drawn as black dashed contours. Thermal particles sourced at the left hand boundary (coming from the interior of the mirror) travel rightwards into a magnetic nozzle. The mirror force and ambipolar potential determine the orbits of incoming particles. Both electrons and ions may be reflected by the mirror force before reaching the throat (as would be expected for mirror confinement). Electrons may also be reflected by the ambipolar potential after crossing the magnetic throat. Particles travelling at sufficiently low pitch angle can travel through the throat to the right hand boundary, an absorbing wall. In regions where the electric and mirror forces are opposed to each other, some particles may become trapped. This occurs to the left of the throat for ions and to the right of the throat for electrons.}
	\label{fig:mirrorsetup}
\end{figure}

We assume that the central region of the magnetic mirror is populated by a thermalized plasma with a Maxwellian distribution for both ions and electrons. This constitutes the upstream condition. Thermal particles traveling along a field line out towards the expander are influenced by an increasing magnetic field strength $B$ and an ambipolar potential drop $\phi_{||}$\cite{yushmanov1966}, and their orbits are fully determined by these profiles in the adiabatic limit. We will see that the form of the distribution function at various points along the flux tube is mediated by qualitatively different guiding center orbits. The primary orbit types are illustrated in Figure \ref{fig:mirrorsetup}. 

The ambipolar electric field is primarily oriented in the direction from the center of the mirror outward through the expander, while the magnetic field takes a maximal value in the expander's throat. The mirror force will accelerate particles away from the throat, while the ambipolar potential accelerates ions rightward towards the device wall and electrons leftward towards the center of the mirror. This divides types of electron and ion orbits according to which side of the throat they are on. Sufficiently energetic particles (ions or electrons) with low pitch angles will be able to travel through the throat and all the way to the absorbing wall without reflecting. Electrons with low pitch angles may be able to traverse the throat, yet still be reflected by the ambipolar potential before reaching the absorbing wall. Particles with pitch angles nearer the perpendicular plane will be more strongly effected by the magnetic mirror force, and may be reflected back towards the center of the mirror before traversing the throat (as one would hope for containment). To the left of throat, since the electric field pushes ions right and the mirror force pushes ions left, trapped ion orbits can exist within the domain considered. Indeed, the reflecting particles are trapped in the magnetic mirror, but not within the local domain considered here. Conversely, trapped electron orbits are possible on the right side of the throat, where the electric and mirror forces are in opposition. Determining which elements of phase space correspond to specific orbit types is key to understanding the particle distribution functions.  

\begin{figure}
	\centering
    \includegraphics[width=0.7\linewidth]{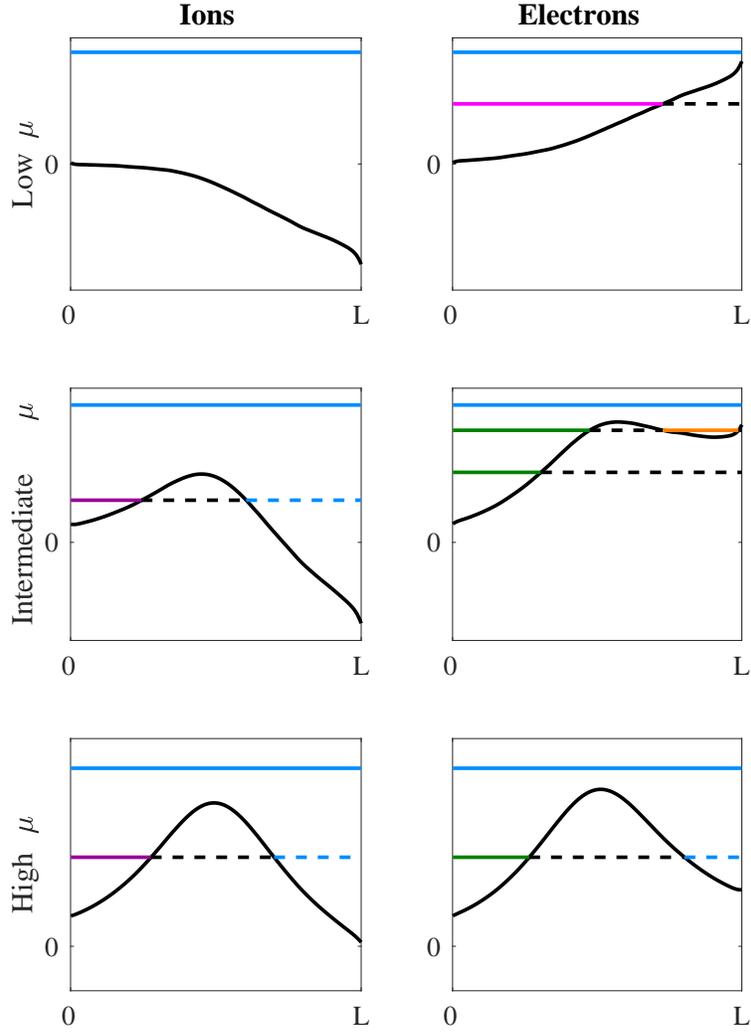}
	\caption{Effective potentials $\mu B + q \phi_{||}$ \cite{yushmanov1966} encountered for ions and electrons at three values of $\mu$. The horizontal levels correspond to different values of $\mathcal E(x_0) + q \phi_{||}(x_0)$. Colors correspond to the orbits drawn in Figure \ref{fig:mirrorsetup}. Dashed lines are blocked but not trapped. The effective potential is dominated by $q \phi_{||}$ for particles with low $\mu$ and $\mu B$ for high $\mu$ particles. In principle, trapped ions can exist, though the shape of $\phi_{||}$ makes these regions very small.}
	\label{fig:EffPot}
\end{figure}

To determine which type of orbit an element of phase space corresponds to, the profiles of $B$ and $\phi_{||}$ along the flux tube must be analyzed. For any magnetic moment $\mu$, the so-called Yushmanov \cite{yushmanov1966} effective potential $\mu B + q \phi_{||}$ describes the parallel dynamics. In other words, the parallel kinetic energy $\mathcal E_{||}$ at any point $x$ on the flux tube is characterized by $\mathcal E_{||}(x) = \mathcal E (x_0) - \mu B(x) - q (\phi_{||}(x)-\phi_{||}(x_0))$, where $\mathcal E$ is the total kinetic energy and $x_0$ is some initial point of consideration on the flux tube. Turning points $x_t$ can be found where $\mathcal E_{||}(x_t) = 0$. To determine whether an orbit will have a turning point to either side of $x_0$, only the maximum value of $\mu B + q \phi_{||} = U_{max\pm}$ to either side is necessary. If $\mathcal E (x_0)  + q \phi_{||}(x_0) < U_{max\pm}$, the particle will eventually reflect if travelling in the corresponding direction; otherwise, the particle can reach the boundary of the considered domain in that direction. We consider an element of phase space to be ``blocked" if $\mathcal E (x_0) + q \phi_{||}(x_0) < U_{max-}$, as an orbit cannot be traced all the way back to the upstream source. We consider an orbit to be ``reflected" or ``reflecting" (depending on the sign of $v_{||}$) if $\mathcal E (x_0) + q \phi_{||}(x_0) < U_{max+}$, as an orbit cannot be traced to the absorbing wall, and thus these particles which started an orbit at the source either have reflected or will reflect back towards the central mirror region. 
\FloatBarrier
Figure \ref{fig:EffPot} illustrates the types of orbits that can occur by determining blocked and reflecting regions, with colors used corresponding to the orbits drawn in Figure \ref{fig:mirrorsetup}. We show the effective potential $\mu B + q \phi_{||}$ for both ions and electrons at three selected values of the magnetic moment $\mu$ plotted against the distance $x$ along the midplane fieldline where thermal particles are injected at $x=0$ and $B(L)=B(0)$. Horizontal lines represent different values of $\mathcal E(x_0) + q \phi_{||}(x_0)$, corresponding to the kinetic energy of a particle at $x=0$, where $\phi_{||} = 0$. Depending on the value of $\mu$ selected, the effective potential may take different shapes. Assuming that the variation of $\phi_{||}$ scales as the electron temperature $T_e$ at the source, for $\mu B_\mathrm{max} \ll T_e$ the contribution of $q \phi_{||}$ dominates, and in this simple scenario the effective potential is monotonic (increasing for electrons, decreasing for ions). As such, all ions that are sourced at $x = 0$ will stream freely without encountering any barriers. Electrons of a high enough energy will also pass all the way to the absorbing wall, but lower energy electrons will not be able to pass beyond an effective potential barrier, as illustrated by the dashed black line indicating a region with $\mathcal E_{||} < 0$. As such, this electron will reflect back towards the origin. 

For those particles with a very large magnetic moment $\mu B_\mathrm{min} \gg T_e$, the effective potential will be dominated by $\mu B$. Such a case does not distinguish between ions and electrons. Both can pass unimpeded if they start at a high enough energy, and will reflect at a lower energy, not being able to enter the region of negative $\mathcal E_{||}$ corresponding to the peak in $B$. If such a particle were to start on the right side of the barrier, it would freely stream to the right boundary, though such a particle is not directly connected to the source. This is represented by the light blue dashed lines, corresponding to physical orbits that will not be filled by the source. 

In the intermediate $\mu$ case, both terms in the effective potential can be important. For the electrons, this means that there are regions to the right of the throat where a local well in the effective potential may form. For low enough and high enough energy electrons, reflecting and passing orbits still exist; however, at a particular energy, trapped particles may be found. A trapped orbit is shown by the orange portion of the line. Starting from this point, an electron is both blocked and reflected, and will remain in this region, though the region is not causally connected to the source. However, the trapped region may be filled through a slower process such as an effective scattering or development of the potential well, as particles that reach a trapped state through one of these processes will remain in the trapped region and not stream away to the wall. Of course, trapped particles may be scattered out of the trapped region if such a process is present; as such, in the presence of an effective scattering process some equilibrium trapped distribution would be formed balancing particles scattering into and out of trapped trajectories. In principle, for some values of $\mu$ ions may also have trapped regions (occurring on the left side of the throat), but the shape of $\phi_\|$ chosen from simulation results makes these regions considerably smaller. As such, the effective potential for the intermediate and high $\mu$ shown is qualitatively identical. 

When determining how to fill regions of phase space, we look at whether the orbit connects back to the source. Passing particles are neither blocked nor reflected. Due to the assumed boundary conditions in this scenario, only those passing regions of phase space with $v_{||}>0$ will map back to the upstream distribution, while passing regions of phase space with $v_{||}<0$ will be empty (in the upcoming Eq.~\ref{eq:distform} these will be denoted as passing$+$ and passing$-$, respectively). Reflecting particles are also simply mapped to the upstream distribution, while reflected particles will be the mirror image of the reflecting particles about $v_{||}=0$, being further down the same orbits. Blocked regions of phase space are inaccessible, save for when they are also reflected/reflecting. These regions are trapped, and particles that enter these portions of phase space will remain there. Trapped particles are assumed to conserve their total energy $\mathcal E + q \phi_\|$ from the upstream population, and map to the corresponding element of phase space. 
We can thus give a form for the distribution function at any point along the flux tube in terms of the upstream distribution function.

\begin{eqnarray}
\label{eq:distform}
    f(\mathbf{x}, \mathbf{v}) = \begin{cases} \xi_t f_\infty(\mathcal E + q \phi_{||}) & \text{trapped}    \\
    f_\infty(\mathcal{E}+ q \phi_{\|}) & \text{passing}+
    \\f_\infty(\mathcal{E}+ q \phi_{\|}) & \text{reflected}\\
    0 & \text{passing}- \\
    0 & \text{blocked}
    \end{cases}
\end{eqnarray}

Where $f_\infty(\mathcal E)$ is the upstream distribution, assumed hereafter to be an isotropic Maxwellian representing the thermal plasma from the interior of the mirror device, and $\xi_t$ is a phenomenological fraction of the trapped population filled, where $\xi_t =1$ makes the distribution continuous at the trapped-passing boundary and $\xi_t=0$ makes the trapped portion of phase space empty. Collisions, transients, or wave scattering typically allow $\xi_t = 1$ for electrons, though the trapped fraction may be considerably smaller for ions. In this paper, for numerical stability we use $\xi_t = 0.02$ for ions, but the trapped regions of phase space for ions are small enough that results are not particularly sensitive to this choice. Through the model of Eq.~\ref{eq:distform}, the distribution is fully specified by the profiles of $B$ and $\phi_{||}$, and moments of the distribution can easily be computed. 

To obtain useful predictions from the model, we numerically determine the profile of the ambipolar potential $\phi_{||}$ as follows. We start with an assumed magnetic field strength profile $B(x)$. We solve for the potential $\phi_{||}(x)$ using a variation of the gradient descent method. An initial guess $\phi^0_{||}$ for the profile $\phi_{||}(x)$ is made ($\phi_{||}(x)=0$ or a linear profile work in practice), and a sequence of improved solutions $\phi^k_{||}(x)$ is then computed through iteration. For each point in the domain, the distributions for electrons and ions are computed based on Eq.~\ref{eq:distform} using the profiles $B(x)$ and $\phi^k_{||}(x)$. Note that the distributions depend on the global profiles of $B(x)$ and $\phi_{||}(x)$ through their values at turning points of particles of different total energy $\mathcal E$ and magnetic moment $\mu$. From the particle distributions, we integrate moments to calculate the electron and ion densities ($n^k_e$ and $n^k_i$) and parallel particle fluxes ($\Gamma^k_e$ and $\Gamma^k_i$) based on $B(x)$ and $\phi^k_{||}(x)$. Skovorodin \cite{Skovorodin2019InfluenceTrap} uses a similar iterative method with different update equations. We iterate to find the consistent limiting potential profile that makes $n_i(x) - n_e(x) = 0$ by setting
\begin{eqnarray}
\phi^{k+1}_{||}(x) &=& \phi^k_{||}(x) + \lambda [\delta \phi^k_{||}(x)]\\
\delta \phi^k_{||}(x) &=& \frac{T_e T_i}{e(T_e+T_i)} \ln \left(\frac{n^k_i(x)}{n^k_e(x)}\right) 
\label{eq:iter}    
\end{eqnarray}
where the form of $\delta \phi^k_{||}(x)$ comes from assuming a Boltzmann-like response of both the electrons and ions to variations in the potential, and a step size of $\lambda=0.67$ typically gives convergence to less than $1\%$ in 10 to 20 iterations. In reality, a Boltzmann response is a crude approximation, which is particularly poor for the ions in the expander. The important features of $\delta \phi^k_{||}(x)$ in Eq.~\ref{eq:iter} are that it gives relatively rapid convergence and that reaching the quasineutrality condition leads to $\delta \phi^k_{||}(x) = 0$. 

During the iteration process, the boundary value of the potential $\phi_w = \phi_{||}(x_w)$, where $x_w$ is the coordinate of the absorbing wall, is likewise adjusted to ensure also that $\Gamma_e = \Gamma_i$. The boundary value $\phi_w$ in general causes a discontinuity in $\phi_\|$ at the wall, which can represent a potential drop across a Debye-scale sheath where quasineutrality is violated \cite{Skovorodin2019InfluenceTrap}. This allows us to capture the effects of the Debye sheath, which is present in a physical device, without needing to resolve it with a more detailed model. In real devices, the potential jump $\Delta \phi$ across the sheath plays a role in regulating the transport of secondary electrons emitted at the wall \cite{ryutov2005,Skovorodin2019SuppressionTrap}. Importantly, the sheath potential jump in expanders with a large magnetic expansion ratio is much less than the electron temperature, $e\Delta \phi/T_e < 1$. Our model agrees with this experimentally verified \cite{Soldatkina2017InfluenceTrap} result.

While equal particle fluxes $\Gamma_e = \Gamma_i$ are appropriate for a quasineutral steady-state GDT, we note that a residual net electrical current may flow across the PIC simulations we present below. For comparison of the drift kinetic model with the kinetic simulations we modify the flux constraint $\Gamma_e=\Gamma_i$ for the model to $\Gamma_e=4\Gamma_i$ as to match the conditions that develop in the kinetic simulation. Although these conditions are somewhat different than those of a physical GDT, this numerical scenario does provide a rigorous test of the drift-kinetic model against the self-consistent kinetic simulation.


\FloatBarrier
\section{Profile of the ambipolar potential and model profiles for a Deuterium device}\label{sec:potential}
\FloatBarrier
In this section, we analyze the predictions of the model of Section~\ref{sec:model} for a device with a high mirror ratio. Many of the qualitative features are generic, and they depend only weakly on the mirror ratio (for $B_m/B_0\gtrsim 15$) and plasma parameters. Here, we consider a case with a mirror ratio of $B_m/B_0=30$, an ion-to-electron mass ratio for a deuterium plasma of $m_i/m_e = 3672$, and equal ion and electron temperatures $T_e=T_i$. The model of Sec.~\ref{sec:model} can easily be applied to this regime relevant to fusion experiments, though it is not feasible to directly simulate these parameters with a fully kinetic code. 

\begin{figure}
	\centering
    \includegraphics[width=0.7\linewidth]{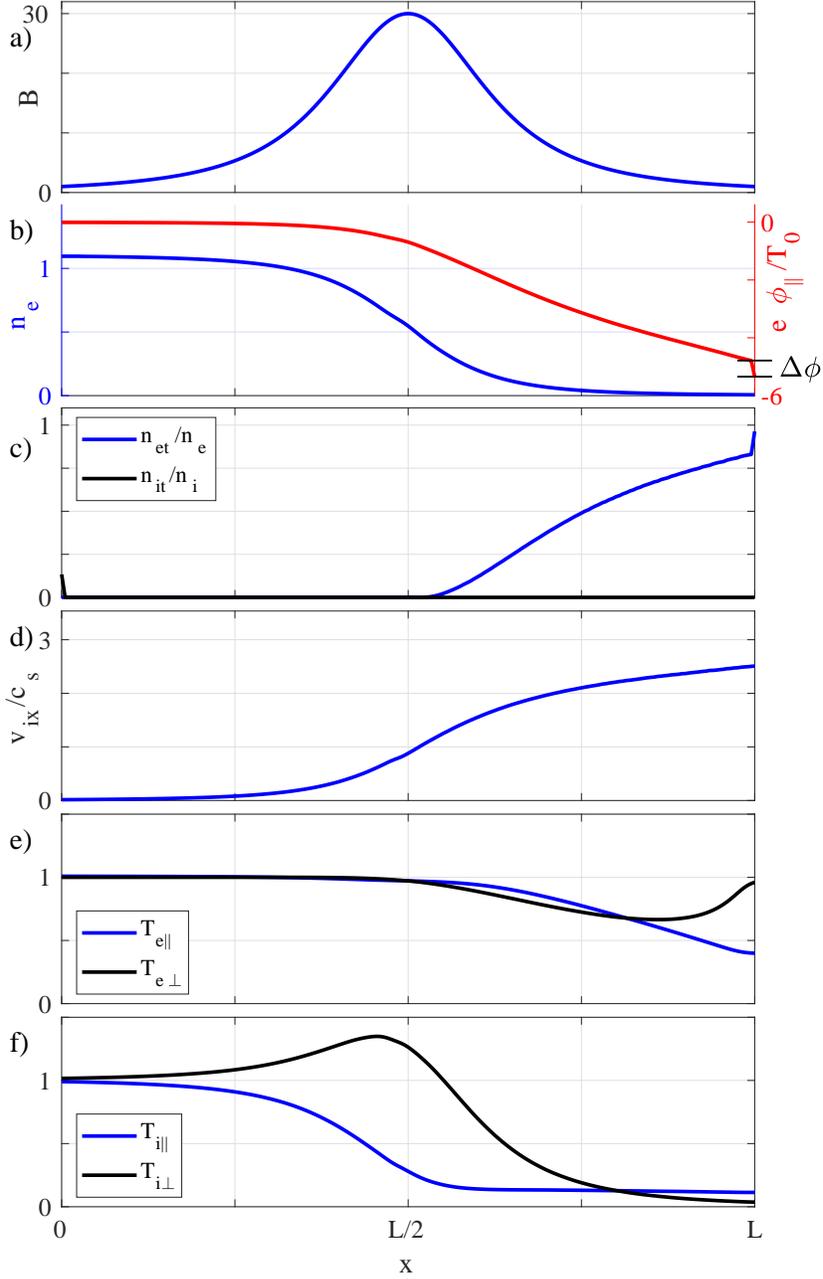}
	\caption{ a) Assumed magnetic field strength profile. Theoretical plots of b) electron density and the parallel potential (where the sheath potential can be seen as the jump in the potential at the absorbing wall, marked on the right), c) the fraction of density corresponding to trapped particles $n_t/n$, d) ion bulk flow velocity along the direction of the magnetic field, normalized to the ion sound speed $c_s = \sqrt{(T_e+3 T_i)/m_i}$, e) electron temperatures, and f) ion temperatures in a deuterium plasma with mirror ratio $B_m/B_0=30$ along the central axis.}
	\label{fig:DeutPlots}
\end{figure}

We assume a magnetic field strength profile as found along the symmetry axis of a single solenoid coil, $B(x) = B_m/(1 + (x-L/2)^2)^{3/2}$ with $L/2 = \sqrt{(B_m/B_0)^{2/3}-1}$, such that $B(0)=B(L)=B_0$, plotted in Figure~\ref{fig:DeutPlots}a). The iterative solution method of Section~\ref{sec:model} is then used to determine the ambipolar electrostatic potential profile $\phi_{||}$ consistent with quasineutrality, equal electron and ion fluxes, and the upstream thermal boundary condition. This ambipolar potential $\phi_{||}$ is plotted on the right axis of Figure~\ref{fig:DeutPlots}b). Notably, the total potential drop across the device is $e\Delta \phi_{||} \approx 5 T_e$, creating a natural thermal barrier for electrons exiting the expander. This is a result of the flux matching conditions between ions and electrons, as analyzed in the solar wind by Boldyrev et al. \cite{Boldyrev2020ElectronWind}. The discontinuity in $\phi_{||}$ to the right wall is the sheath potential, which is on the order of $T_e/2e$. The ambipolar electric field accelerates ions out of the magnetic trap, and the bulk ion flow velocity in Figure~\ref{fig:DeutPlots}d) becomes superthermal past the magnetic throat.

The electron density profile responds to $\phi_{||}$ and is considerably depleted by the right wall. We consider temperature moments of the non-Maxwellian particle distributions given by $T_{||} = m/n\int f({\bf{v}})(v_{||}-u_{||})^2 d^3v$ and $T_{\perp} = m/2n\int f({\bf{v}}){v_{\perp}}^2 d^3v$, where $u_{||}$ is the bulk parallel fluid flow speed for the species. The electrons to the left of the magnetic throat (the region of strong $B$ field) are essentially isothermal, as in the Boltzmann limit of working against an electric field appropriate for particles with fast thermal transit times relative to other time scales in the system. To the right of the throat, both $T_{e||}$ and $T_{e\perp}$ as plotted in Figure~\ref{fig:DeutPlots}e) initially dip, though the perpendicular electron temperature returns to approximately its original value at the absorbing wall. The detailed electron temperature profile to the right of the peak in $B$ is relatively more sensitive to the population of locally trapped electrons represented in Equation~\ref{eq:distform}. This trapped electron population could in practice result from weak collisions, transients initiated during plasma start-up, or scattering by waves and instabilities. For our plots, we use a trapped filling fraction of $\xi_t=1$, which tends to result in the weakest temperature gradients and places a rough lower bound on the changes in $T_e$ in the expander exhaust. Figure~\ref{fig:DeutPlots}d) plots the fraction of electrons and ions trapped $n_t/n$ over the course of the device. Ion trapping is negligible almost everywhere, while the majority of electrons deep in the expander are trapped.

The ion temperature profiles, plotted in Figure~\ref{fig:DeutPlots}f), are less flat to the left of the throat, with $T_{i||}$ decreasing and $T_{i\perp}$ increasing as $B$ increases, consistent with the conservation of energy and magnetic moment in a region where $\phi_{||}$ is relatively flat. To the right of the throat, $T_{i\perp}$ falls with the decreasing magnetic field strength while $T_{i||}$ is approximately isothermal with the temperature reduced from its source value. Ions to the right of the throat are essentially beaming and have no reflected component. This cold beam velocity distribution reduces the spread of the parallel velocity and explains the lower $T_{i||}$.

\FloatBarrier
\section{Particle-in-cell Simulations}\label{sec:basesim}
We compare our drift-kinetic model to first-principles particle-in-cell (PIC) simulations of plasma exiting a magnetic trap. The simulations solve the collisionless Maxwell-Vlasov equations using the fully relativistic and electromagnetic PIC code VPIC \cite{Bowers2008}. A 2-D slab magnetic nozzle geometry is produced in the $x-z$ simulation plane by a set of external coils \cite{Guo2012a} with a mirror ratio along the central axis ($z=0$) of $B_m/B_0=5$. The domain is initially filled with a uniform plasma of density $n_0$ sampled by 800 numerical particles per species per cell. The grid is of size $L_x\times L_z = 200~d_e\times100~d_e$ (where $d_e = \sqrt{\epsilon_0m_ec^2/n_0e^2}$ is the electron inertial length) resolved by $2016\times1008$ grid points, and a time step of $dt \approx 0.07/\omega_{pe}$ is used. The electrons and ions have equal temperature $T_0$, which is chosen to give an electron beta of $\beta_e = 2n_0T_0/\mu_0B_m^2=0.03$ in the high-field region at the center of the mirror throat. A reduced ion-to-electron mass ratio of $m_i/m_e=400$ is used, and the ratio of the plasma frequency to the electron cyclotron frequency at the center of the domain is $\omega_{pe}/\omega_{ce}=1.22$.

The boundaries of our simulations use a local magnetic flux-freezing boundary condition that sets the tangential (to the boundary surface) components of the electric field to zero. The boundaries absorb particles, and the absorbed charge is accumulated as a local bound charge. The bound charge at each cell allows a normal electric field to develop at the boundary surface, which forms a Debye-scale sheath potential. To represent a thermal plasma source on the left ($x=0$) boundary, we inject a Maxwellian flux of electrons and ions based on distributions at fixed density $n_0$ and temperature $T_0$, as indicated in Figure \ref{fig:mirrorsetup}. After $\sim 7$ ion thermal crossing times ($L_x/v_{thi} = L_x/\sqrt{T_0/m_i}$), the system settles into a quasi-steady state. We compare the fields (averaged over 200 simulation time steps to smooth over statistical particle noise) and electron and ion profiles from this late-time quasi-steady state to the drift kinetic model.

\begin{figure}
	\centering
    \includegraphics[width=1.0\linewidth]{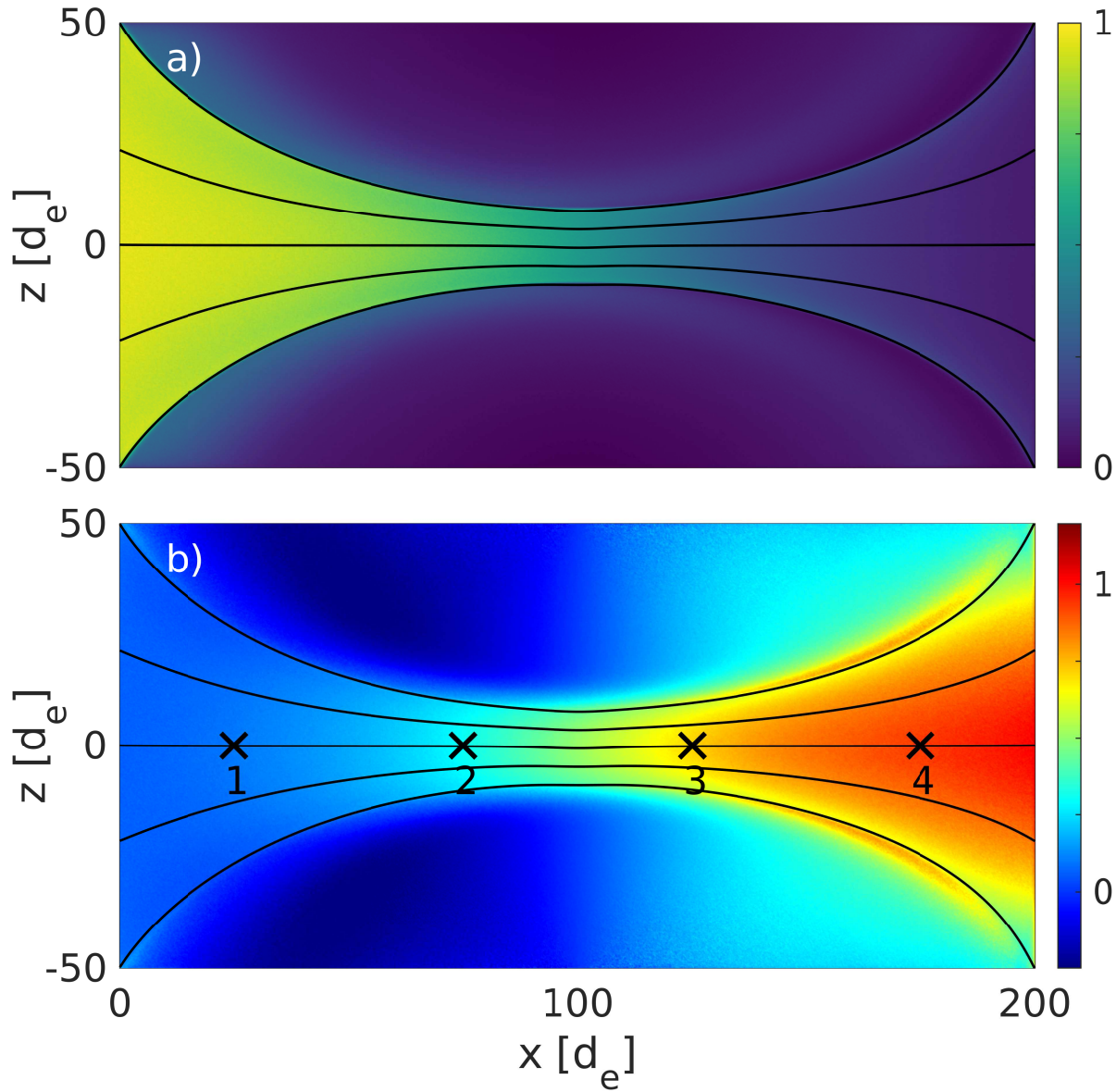}
	\caption{Profiles of a) ion density and b) ion bulk velocity (normalized to the ion sound speed $c_s = \sqrt{(T_e+3 T_i)/m_i}$) in the VPIC simulation. Selected field lines are shown in black.}
	\label{fig:NBpcolor}
\end{figure}

We begin by showing contours of the ion density and the $x$-component of the ion bulk velocity in Figure \ref{fig:NBpcolor}. The bulk of the ions are confined to the interior of the mirror and do not cross the throat. Those that do make it across the throat travel right towards the absorbing wall supersonically.

In order to compare the model and simulation properly, profiles for $B$ and $\phi_{||}$ must be specified. The magnetic field at the midplane is dominated by the field of the external coil. While the profile of the magnetic field is quite well-known, there is a degree of uncertainty in the profile of $\phi_{||}$, largely due to the inherent noisiness of PIC electric fields. The time-averaging applied to the fields helps to reduce this uncertainty, but does not eliminate it altogether. As such, we apply the iterative approach detailed in Eqs.~\ref{eq:iter} to solve for the model potential with one key alteration. The local field and particle boundary conditions in simulation allows for a net global current through the simulation domain, which we would not expect to see in a physical device. The fixed Maxwellian particle source at the left boundary injects electrons with a flux $v_{the}/v_{thi} = 20$ times greater than that of the ions. Although an electric potential develops across the simulation, there remains a residual net electrical current. The electron and ion fluxes are observed to self-consistently adjust to a ratio of $\Gamma_e / \Gamma_i \approx 4$ in the simulation (rather than a ratio of 1 expected in a real experiment). To match this, we adjust the boundary value of the potential $\phi_w$ in the drift-kinetic model such that $\Gamma_e / \Gamma_i = 4$. Especially for the ions, this causes only moderate modifications to the predicted temperature and density profiles.

\begin{figure}
	\centering
    \includegraphics[width=0.8\linewidth]{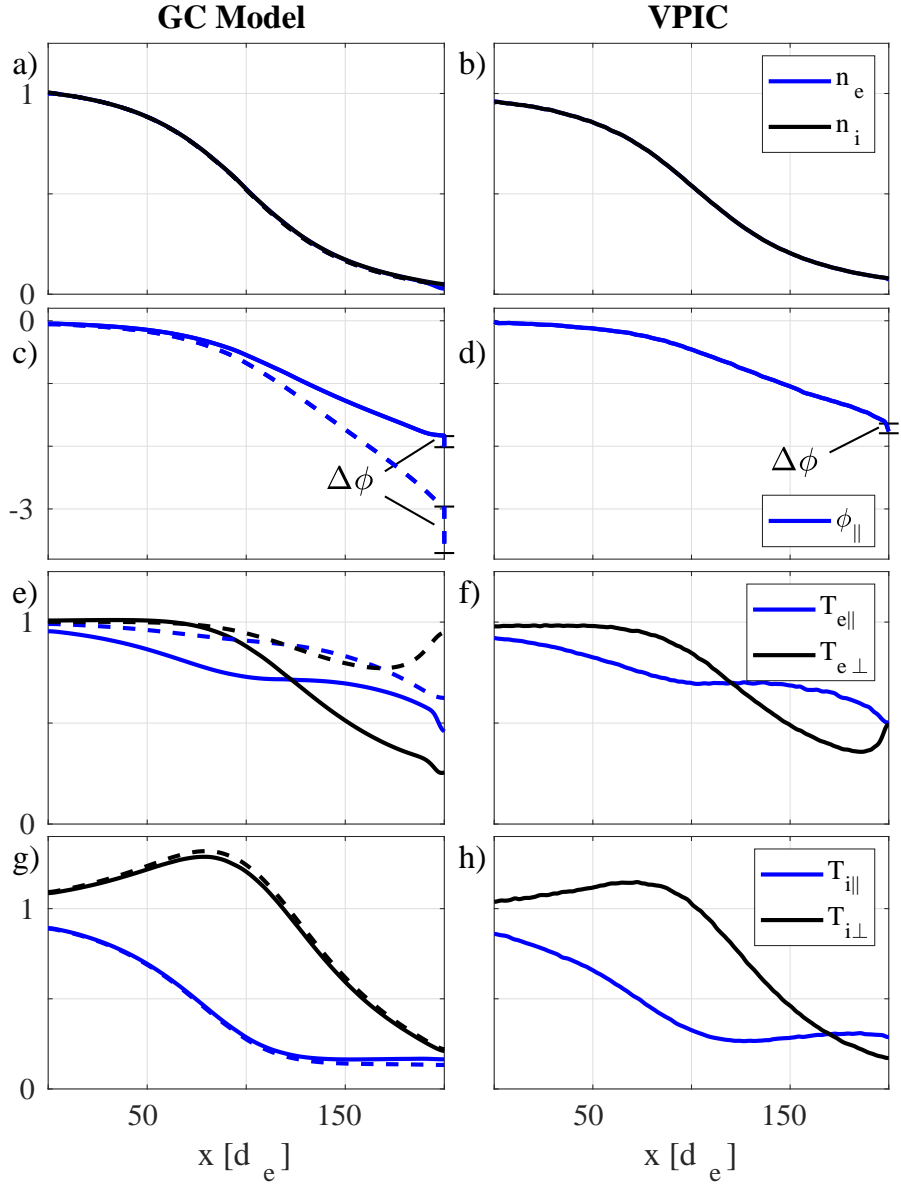}
	\caption{Profiles of a-b) electron and ion density, c-d) ambipolar potential with the sheath potential $\Delta \phi$ marked on the right, e-f) electron temperatures, and g-h) ion temperatures for the guiding center model and VPIC simulation. Solid lines represent the flux condition that matches the simulation conditions, $\Gamma_e=4 \Gamma_i$. Dashed lines represent the more physical condition of $\Gamma_e= \Gamma_i$. The ions are relatively insensitive to this choice, though electrons would be expected to experience less temperature variation in a physical device than in the simulation.}
	\label{fig:NBplots}
\end{figure}

 With the procedure for adjusting the potential profile established, we compare the profiles of ion and electron fluid quantities along the central field line at $z=0$ observed in the simulation to the predictions of the drift-kinetic model. This comparison is shown in Figure \ref{fig:NBplots}. We present model curves both for the observed condition in the simulation that $\Gamma_e=4 \Gamma_i$ (solid curves) and the more physical condition that $\Gamma_e=\Gamma_i$ (dashed curves). The density profiles, being an input to the guiding center model, match. While the potential that develops in the simulation is well-matched by the model using the simulation flux condition, a more physical system would be expected to have both a larger potential develop across the device and a larger potential jump at the sheath. When looking at the solid curves, the electron parallel and perpendicular temperature profiles are a good match, with $T_{e\perp}>T_{e||}$ up to the throat, and $T_{e\perp}<T_{e||}$ in the exhaust. The ion temperature profiles are also well-predicted, with $T_{i\perp}>T_{i||}$ for nearly the whole device. We can see that the ion temperature profiles are not particularly sensitive to the flux condition, while the electron temperature profiles for a physical device would be expected to exhibit significantly less variation. The model is flexible enough to handle boundary conditions appropriate to both the simulation and a physical device, and the more physical boundary condition is indeed significantly easier to implement in the model than in the fully-kinetic simulation.   
 
\begin{figure}
	\centering
    \includegraphics[width=1.0\linewidth]{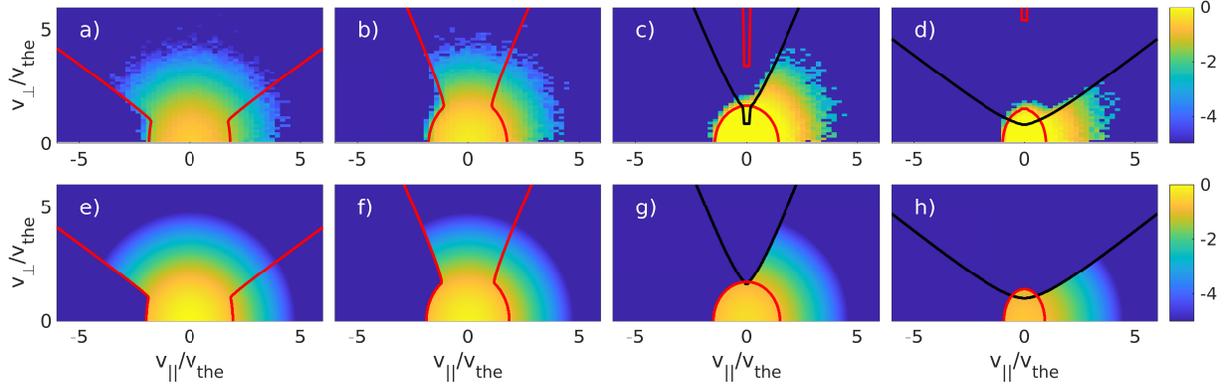}
	\caption{a-d) From left to right, electron distribution functions obtained from the VPIC simulation at the points marked 1-4 in Figure \ref{fig:NBpcolor}. e-h) Model electron distributions at the corresponding points along the central field line. Red contours represent the reflecting boundary and black contours represent the blocking boundary, which determine the shape of the distribution function.}
	\label{fig:edists}
\end{figure}

 To fully verify the drift-kinetic model, electron and ion distributions should be compared between simulation and theory. We begin with the electrons in Figure \ref{fig:edists}. Here the top row consists of electron distributions measured in the simulation at equally-spaced points marked in Figure \ref{fig:NBpcolor}, while the bottom row shows the distributions predicted at the equivalent points in the drift-kinetic model. On each distribution, we mark two contours (if they fall within the displayed region of velocity space) corresponding to the reflecting (red) and blocking (black) boundaries as described in Section \ref{sec:model}. These contours differ slightly between the simulation and theory, as the iterative potential is used in the model but the raw measured potential is used in the simulation. The shape of the model distributions are, of course, determined by these boundaries. We see that the simulation's distributions also display sharp features aligned with the blocked and reflecting boundaries. Blocked regions of velocity space do not have particles unless they are also reflecting (and thus trapped). Reflecting particles are seen going both directions along the field line. The blocked/reflecting dynamics fully explain the form of the electron distribution function. Trapped electrons are seen in the distribution. In theory, these regions are supposed to be filled over time by weak collisional processes; however, the simulation employed here does not include a collisional model. As such, the trapped portions of the distribution are likely sourced from the original background population, which has largely been cleared from the simulation domain, rather than the injected Maxwellian population. 
 
\begin{figure}
	\centering
    \includegraphics[width=1.0\linewidth]{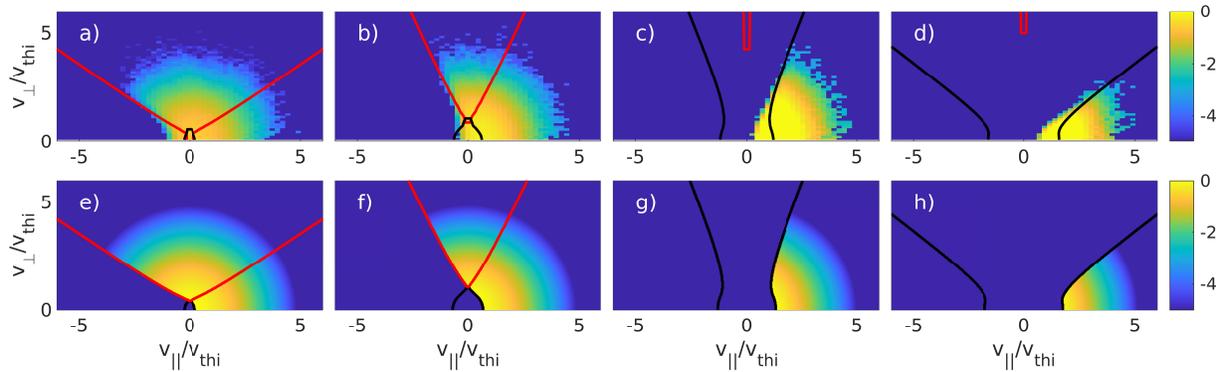}
	\caption{a-d) From left to right, ion distribution functions obtained from the VPIC simulation at the points marked 1-4 in Figure \ref{fig:NBpcolor}. e-h) Model ion distributions at the corresponding points along the central field line. Red contours represent the reflecting boundary and black contours represent the blocking boundary, which determine the shape of the distribution function.}
	\label{fig:idists}
\end{figure}

We now examine the ion velocity-space distribution functions in Figure \ref{fig:idists}. Again, the top row corresponds to the simulation and the bottom to the model at the same points as the electron distributions were shown. The shapes of the blocking and reflecting contours have altered slightly from their form for the electrons, owing to the opposite charge of the ions. Again, the model generally matches the form of the distribution function, though there are some larger discrepancies than there were for the electrons, particularly for those particles with the smallest $\mu$. This is likely due to uncertainty in the profile of $\phi_{||}$, which has its largest effect on low $\mu$ particles.  

\FloatBarrier
\section{Sloshing ion beams}\label{sec:slosh}
A large portion of GDT heating is usually done with neutral beam injection. This introduces a population of fast sloshing ions, which are high-energy and low-collisionality. The sloshing ions have a density profile peaked near the magnetic throats if the velocity pitch angle of the injected beams is chosen near the trapped-passing boundary. The fast ions can amplify fusion reactivities above Maxwellian rates \cite{maximov2004}, help protect against instabilities \cite{coensgen1975}, and may have a plugging effect that reduces end losses \cite{kotelnikov1985}. In this section, we analyze the effect of these fast ions on the parallel electron confinement and development of the parallel ambipolar potential. We demonstrate that the drift-kinetic model of Sec.~\ref{sec:model} is sufficiently general to appropriately model this regime as well. 

To explore the effect of sloshing ion beams, we employ another VPIC simulation, similar to that of Section \ref{sec:basesim}, but with an additional beam ion population added in. The beam is injected on the left hand border at a density of $n_{b0} = 0.03 n_0$ with kinetic energy $20 T_{i0}$ at a pitch angle of $\arcsin(1/\sqrt{10}) \approx 18.4^\circ$, selected so that the fast ions reflect near the center of the magnetic throat. The density of thermal ions injected on the left boundary is reduced to $0.97 n_0$ to maintain neutral charge density. Because the peak of the fast ion density remains low relative to the central thermal ion density, this simulation falls in a regime with only moderate modifications compared to GDT with only a single thermal ion population \cite{kotelnikov1985}. 

\begin{figure}
	\centering
    \includegraphics[width=0.7\linewidth]{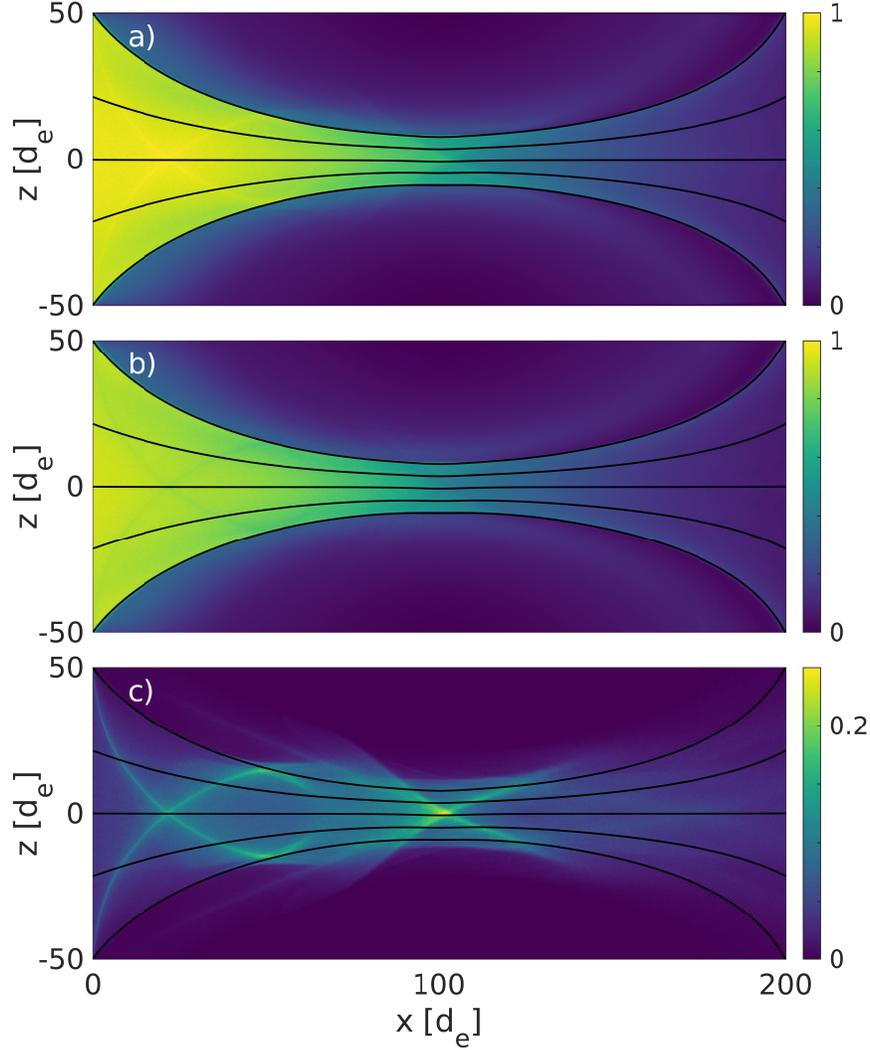}
	\caption{Number densities of a) electrons, b) ions of thermal origin, and c) beam ions for the mirror setup with sloshing ion beams at quasi-steady state.}
	\label{fig:BeamDens}
\end{figure}

Results from the simulation are plotted in Figure \ref{fig:BeamDens}, which shows the densities of the three species after the simulation has a reached a quasi-steady state (the same time as was evaluated in Figure \ref{fig:NBpcolor} for the first simulation). In the broad sense, the electron and ion density profiles shown in Figure \ref{fig:BeamDens}a-b) are similar to the density profile of the original simulation shown in Figure \ref{fig:NBpcolor}a); however, the electron density is no longer identical to density of ions originating from a thermal source. Quasineutrality now requires that $n_e = n_i + n_b$, and the beam density can be seen to alter the densities of the other two species. Figure \ref{fig:BeamDens}c) shows the density of beam ions $n_b$ in the simulation. The fast ion density $n_b$ is peaked in the high-field throat, with additional smaller peaks focused along along caustic lines related to the single-particle trajectories in the curved magnetic geometry in a regime where the beam ion Larmor radius is an appreciable fraction of the domain size. While faint on the scale of the background density, it can be seen that $n_e$ is enhanced and $n_i$ is diminished wherever $n_b$ is peaked. 

\begin{figure}
	\centering
    \includegraphics[width=1.0\linewidth]{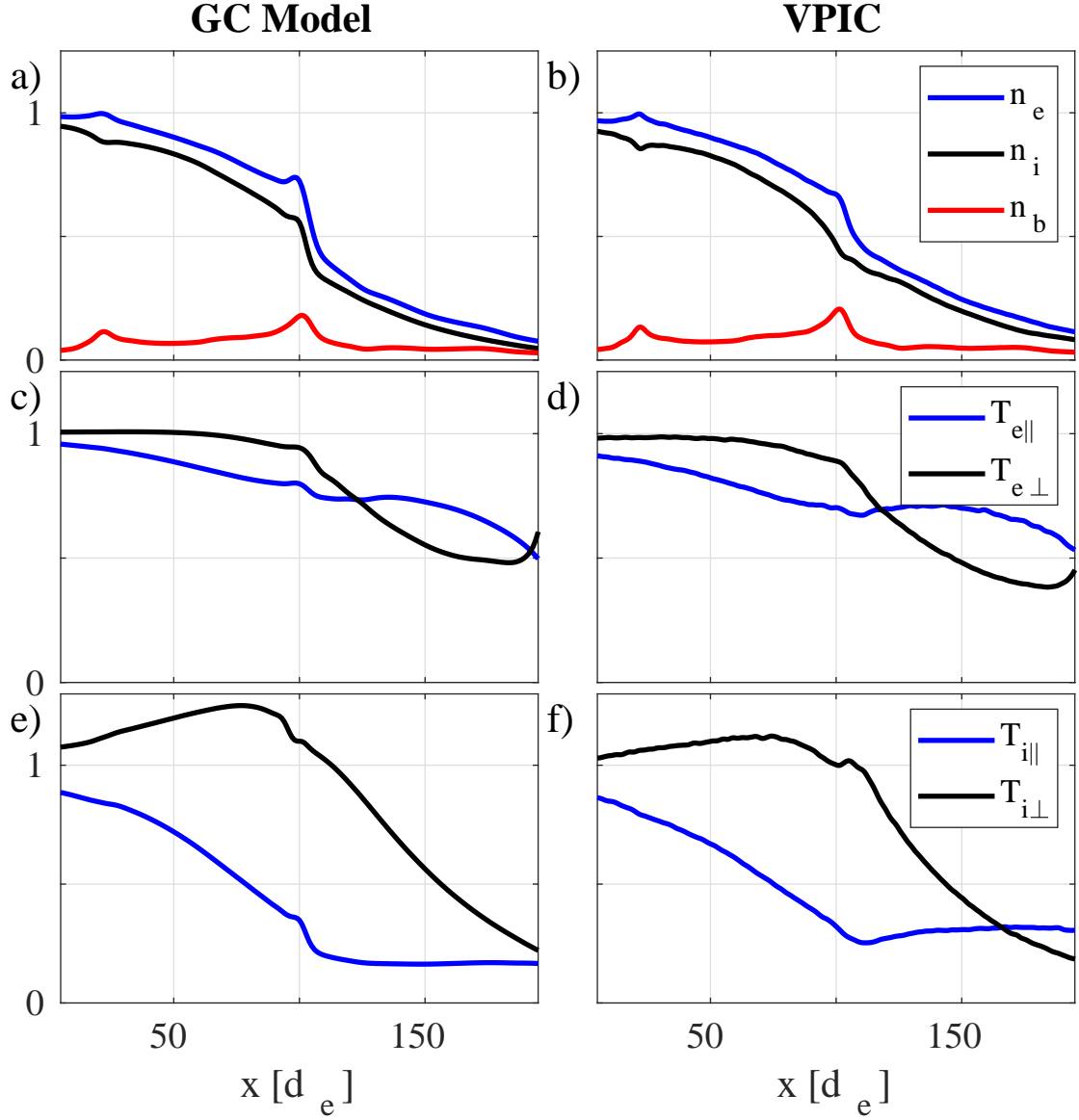}
	\caption{Profiles of a-b) electron, ion, and beam density, c-d) electron temperatures, and e-f) ion temperatures for the guiding center model and the VPIC simulation with a sloshing ion beam. }
	\label{fig:Bplots}
\end{figure}

\begin{figure}
	\centering
    \includegraphics[width=1.0\linewidth]{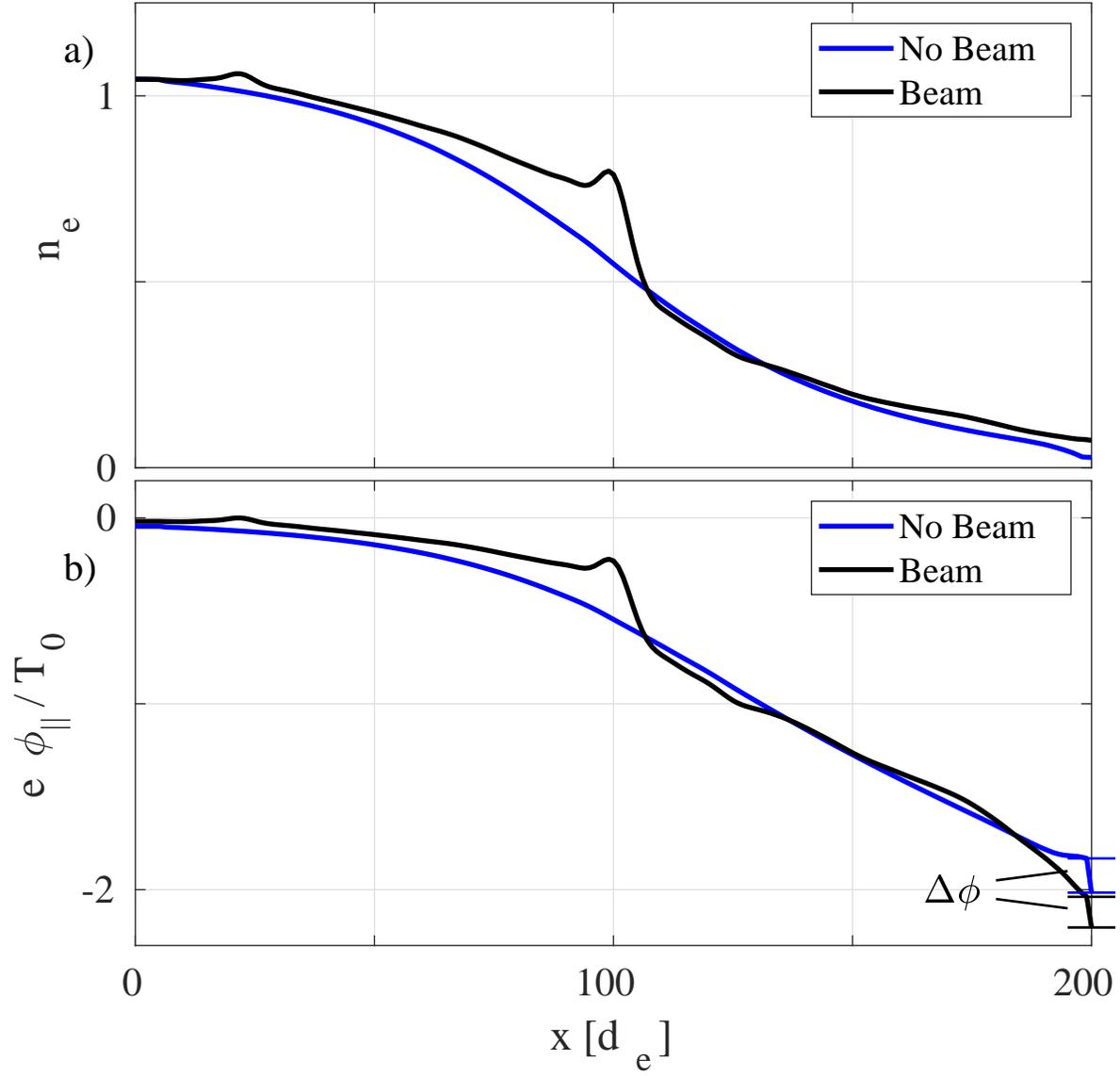}
	\caption{a) The electron density and b) The ambipolar potential $\phi_{||}$ normalized by the initial electron temperature used in the model corresponding to the two simulations presented in this paper. The addition of a sloshing ion beam adds new features to the profile of $\phi_{||}$ where the beam density $n_b$ is enhanced. The sheath potential $\Delta \phi$, as marked at the right wall, does not change much.}
	\label{fig:phicomp}
\end{figure}

One result of adding the sloshing ion beam is that there is a more pronounced drop off in density in the throat to the right of the beam's convergence point. This is clear in the simulation density profiles along the central field line shown in Figure \ref{fig:Bplots}b), and it corresponds to a locally sharp drop in $\phi_{||}$ at this location. As can be seen in Figure \ref{fig:phicomp}b), the overall potential drop across the device is similar for the scenarios with and without sloshing ion beams, but additional local structure has been given to the profile of $\phi_{||}$ where $n_b$ is enhanced. In particular, the enhanced beam density in the throat corresponds to a local bump in $\phi_{||}$, followed by a steep drop off. This makes the potential nonmonotonic, and it allows for additional regions of local trapping and more complicated orbit dynamics, which will not be analyzed in detail here.

We alter the iterative scheme slightly to maintain quasineutrality with a prescribed beam density. We assume, in addition to the magnetic field strength profile $B(x)$, a background ion density $n_b(x)$, which may represent a given profile of fast ions in a GDT that are not solved for within this model. With a Boltzmann response from both the ions and electrons, the quasineutrality condition can then be expressed as $n_b(x) + n_i^k (x) \exp{(-e \delta \phi_{||}^k(x) /T_i)}- n_e^k(x) \exp{(e \delta \phi_{||}^k(x) /T_e)} =0$. This is not generally analytically solvable for $T_i \neq T_e$; however, it is for the case of $T_i = T_e$ which we have used for our simulations. By assuming that the temperature scaling reduces to that of Eq.~3 as $n_b(x) \to 0$, we arrive at an approximation for the case where $T_e \neq T_i$. The resulting replacement for Eq.~3 is 

\begin{equation}
\label{eq:quaddP}
    \delta \phi^k_{||}(x) = \frac{2 T_e T_i}{e(T_e+T_i)} \ln \left(\frac{n_b(x)+\sqrt{n_b(x)^2 + 4 n^k_e(x) n^k_i(x)}}{2 n^k_e(x)}\right) 
\end{equation}

We again apply the guiding center model and iterative scheme for $\phi_{||}$, this time also incorporating $n_b$ through Eq.~\ref{eq:quaddP}, to reproduce the VPIC profiles of fluid moments, as shown in Figure \ref{fig:Bplots}, where parallel and perpendicular electron and ion temperature profiles are again predicted by the guiding center model. Additional structure has been added to the temperature profiles where $n_b$ is enhanced in the throat, though the profiles are broadly similar to the profiles without the sloshing ions save for the sharper density drop in the throat.

\FloatBarrier
\section{Discussion and Conclusion}\label{sec:conc}
In this paper, we have shown that a guiding center model can appropriately describe the physics of the expander region of a magnetic mirror device like GDT. The physics of the expander is fundamentally characterized by a drift-kinetic model that maps several classes of orbits to an upstream distribution function through Liouville's theorem. An ambipolar potential develops across the expander, which the model finds by enforcing quasineutrality and ensuring the net current is zero. An iterative scheme efficiently determines $\phi_{||}$, and the resulting ambipolar potential forms a built-in thermal barrier for electrons exiting the expander in a mirror confinement device. A comparison between the model and fully kinetic VPIC simulations revealed that the model properly predicts the profiles of fluid properties along the device's central field line as well as the form of the ion and electron distribution functions in velocity space. Finally, we showed that the model can incorporate an additional component of sloshing ion beams, which help stabilize the GDT device, and appropriately predicts the fluid profiles along the field lines. 

Going forward, the model can be used to predict profiles in mirror devices. The effects of varying the parameters of the sloshing ion beam injection can also be characterized in this framework. While we considered here a case with relatively low fast ion density, GDT experiments have already been performed with peak fast ion densities exceeding the thermal ion density \cite{ivanov2017}. This regime will be studied in future work. An attractive application of the model would be to create a fluid closure that captures the electron kinetic physics encoded in its framework \cite{ohia:2012,Le2016,Wetherton2019}. Such a closure could enable full-device fluid or hybrid (kinetic ion/fluid electron) simulations to incorporate the physics of the ambipolar potential, allowing an accurate representation of the device that is not feasible on large scales through fully kinetic simulation. Eventually, more complicated transport processes related to collisions with the wall and neutrals, as well as kinetic plasma instabilities, will have to be incorporated into the models \cite{ryutov2005}. Several of these transport processes are also expected to operate in the solar wind, and future studies similar to the one presented here may help refine models of the solar wind temperature \cite{Boldyrev2020ElectronWind}, including temperature anisotropy- and heat flux-driven instabilities.

\begin{acknowledgments}
Research presented in this article was supported by the Laboratory Directed Research and Development program of Los Alamos National Laboratory under project 20200587ECR. Simulations used Los Alamos Institutional Computing resources. The work of SB was supported by the NSF grants No. PHY-1707272, PHY-2010098, by the NASA grant No. 80NSSC18K0646, and by the DOE grant No. DESC0018266.

The data that support the findings of this study are available from the corresponding author
upon reasonable request. Alternatively, VPIC is an open source code available at \url{github.com/lanl/vpic}.  Results can be reproduced by running simulations with the same initial conditions as listed in the paper.
\end{acknowledgments}

\appendix

\bibliography{references.bib}

\end{document}